\documentclass[lettersize,journal]{IEEEtran}

\usepackage{graphicx}
\usepackage{cite}
\usepackage{hyperref}
\usepackage{amssymb}
\usepackage{amsfonts}
\usepackage{amsmath}

\DeclareMathOperator*{\argmin}{argmin}

\usepackage{epsfig}
\usepackage{color}
\usepackage{fancybox}
\usepackage{textcomp}
\usepackage{multirow}
\usepackage{setspa ce}
\usepackage{psfrag}
\usepackage[ruled,vlined,linesnumbered]{algorithm2e}
\usepackage{makecell}
\usepackage{mathrsfs}

\usepackage{booktabs}
\usepackage{threeparttable}

\def\BibTeX{{\rm B\kern-.05em{\sc i\kern-.025em b}\kern-.08emT\kern-.1667em\lower.7ex\hbox{E}\kern-.125emX}}
\usepackage{balance}

    \def\Complex{{\rm\rule[.23ex]{.03em}{1.1ex}\kern-.3em{C}}}

    \newcommand{\be}{\begin{equation}} \newcommand{\ee}{\end{equation}}
    \newcommand{\bea}{\begin{eqnarray}} \newcommand{\eea}{\end{eqnarray}}
    \newcommand{\benum}{\begin{enumerate}} \newcommand{\eenum}{\end{enumerate}}

    \newcommand{\qu}{{\bf u}}
    
    \newcommand{\qw}{{\bf w}}

    \newcommand{\qG}{{\bf G}}
    \newcommand{\qH}{{\bf H}}

    \newcommand{\qW}{{\bf W}}

    \newcommand{\qtheta}{{\boldsymbol \theta}}

    \newcommand{\tg}{\tilde{g}}

    \newcommand{\bbC}{{\mathbb C}}

\begin{document}
\title{AI-based CSI Feedback with Digital Twins: Real-World Validation and Insights}

\author{~Tzu-Hao~Huang,~Chao-Kai~Wen,~\IEEEmembership{Fellow,~IEEE}, Shang-Ho Tsai,~\IEEEmembership{Senior Member,~IEEE},~and ~Trung~Q.~Duong,~\IEEEmembership{Fellow,~IEEE}

\thanks{{T.-H.~Huang} is with the Institute of Electrical Control Engineering, National Yang Ming Chiao Tung University, Hsinchu 300, Taiwan, Email: {\rm  peter94135@gmail.com}.}
\thanks{{C.-K.~Wen} is with the Institute of Communications Engineering, National Sun Yat-sen University, Kaohsiung 804, Taiwan, Email: {\rm chaokai.wen@mail.nsysu.edu.tw}.}
\thanks{{S.-H.~Tsai} is with the Department of Electrical Engineering, National Yang Ming Chiao Tung University, Hsinchu 300, Taiwan, Email: {\rm shanghot@mail.nctu.edu.tw}.}
\thanks{T. Q. Duong is with the Faculty of Engineering and Applied Science, Memorial University, St. John's, NL A1C 5S7, Canada, and with the School of Electronics, Electrical Engineering and Computer Science, Queen's University Belfast, Belfast, U.K, Email: tduong@mun.ca.}

}

\maketitle

\begin{abstract}
Deep learning (DL) has shown great potential for enhancing channel state information (CSI) feedback in multiple-input multiple-output (MIMO) communication systems, a subject currently under study by the 3GPP standards body. Digital twins (DTs) have emerged as an effective means to generate site-specific datasets for training DL-based CSI feedback models. However, most existing studies rely solely on simulations, leaving the effectiveness of DTs in reducing DL training costs yet to be validated through realistic experimental setups. This paper addresses this gap by establishing a real-world (RW) environment and corresponding virtual channels using ray tracing with replicated 3D models and accurate antenna properties. We evaluate whether models trained in DT environments can effectively operate in RW scenarios and quantify the benefits of online learning (OL) for performance enhancement. Results show that a dedicated DT remains essential even with OL to achieve satisfactory performance in RW scenarios.
\end{abstract}

 
\section{Introduction}
\IEEEPARstart{E}{nhancing} spectrum efficiency through large-scale multiple-input multiple-output (MIMO) systems, which deploy numerous transmitting antennas at base stations (BSs), is a key strategy in 5G and emerging 6G communication systems \cite{MIMO}. Fully capitalizing on the benefits of large-scale MIMO requires accurate channel state information (CSI) at the BS, as CSI quality directly influences system performance \cite{CSIT}. In frequency division duplex (FDD) systems, downlink CSI is estimated at the user equipment (UE) and fed back to the BS. Current 5G New Radio (NR) systems, such as those employing Type II feedback \cite{3GPP_CSI_Feedback}, rely on codebook-based methods. However, as the number of antennas increases, codebook design becomes inefficient, leading to degraded CSI quality or excessive feedback overhead. 

Recent deep learning (DL) advances have demonstrated strong potential for CSI feedback \cite{CsiNet}. DL-based autoencoders have been applied to compress and reconstruct complete downlink CSI, while subsequent work \cite{EVCsiNet} has introduced implicit feedback mechanisms that focus on feeding back precoding matrices rather than full CSI. This method offers higher accuracy than traditional codebook schemes and is compatible with the 5G NR standard.  
More recent efforts further improve feedback efficiency by incorporating quantization-aware training \cite{CQNet} and entropy coding techniques \cite{DeepCMC}.

However, training DL models requires large datasets that are challenging to obtain in real-world (RW) systems without performance degradation. Digital twins (DTs) have emerged as an efficient solution for generating site-specific datasets for DL model training \cite{DT_FL,DT_MIMO,DigitalTwin_CSI_Feedback,Learnable_Wireless_DT}.
Additionally, recent developments \cite{Geo2ComMap} have demonstrated that low-cost DT construction is feasible using publicly available data and open-source tools, further supporting the practicality of DT-based training.
Nevertheless, discrepancies between DT simulations and RW environments often lead to performance degradation when models trained on DT data are deployed in practice. Integrating online learning (OL) techniques that leverage minimal RW data is essential to improve model performance \cite{DigitalTwin_CSI_Feedback}. To date, most research has relied on simulations, and a realistic experimental setup is needed to validate the effectiveness of DTs in reducing DL training costs. 

This paper addresses this gap by establishing an RW environment and corresponding virtual channels using ray tracing with replicated 3D models and accurate antenna properties. We construct a comprehensive dataset that includes RW measurements and virtual channel data derived from these measurements. This dataset is used to train a DL-based autoencoder for implicit CSI feedback. To bridge the gap between virtual and real channels, OL is employed to fine-tune the DL models with a minimal amount of RW data. Through detailed experiments, we compare the communication performance of different feedback schemes and address the following questions: \emph{Can a model trained in a DT environment perform effectively in RW scenarios?} \emph{How does OL improve performance?} 

\begin{figure*}
    \begin{center}
        \resizebox{7in}{!}{
            \includegraphics*{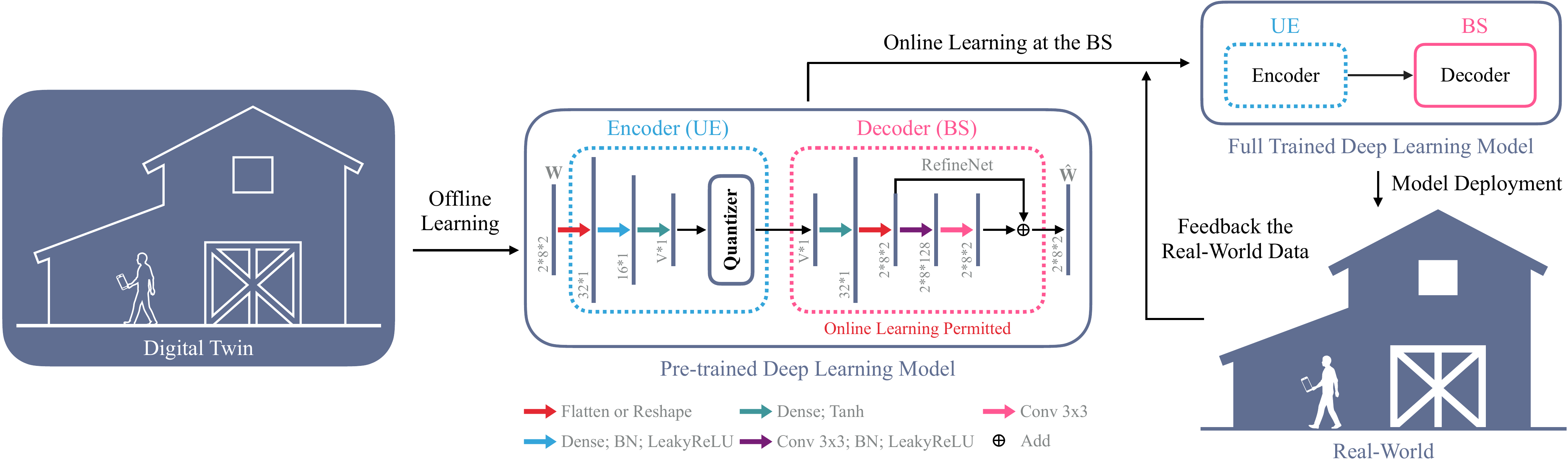} }
        \caption{Neural network architecture and an overview of the efficient AI-driven CSI feedback training approach integrating DTs and OL.}\label{fig:Key_Point_NN}
        \vspace{-0.35cm}
    \end{center}
\end{figure*}

\section{System Model and Design with DT}

\subsection{Communication System Model}
We consider a single-user MIMO system with $N_\text{t}$ transmit antennas at the BS and $N_\text{r}$ receive antennas at the UE. The system employs an orthogonal frequency-division multiplexing (OFDM) waveform with $N_\text{c}$ subcarriers. 
In the 5G New Radio (NR) system, the UE acquires the downlink CSI, computes $N_\text{s}$ eigenvectors (or spatial precoders), and feeds them back to the BS via uplink transmission. The $s$-th eigenvector, $\qw_s \in \mathbb{C}^{N_\text{t}}$, normalized such that $\|\qw_s\|_2 = 1$, is obtained by eigenvalue decomposition:
\begin{equation}\label{eq:eig_de}
{\left( \frac{1}{N_\text{c}} \sum_{n=1}^{N_\text{c}} \qH_n^\mathrm H{\qH}_n\right)} \qw_s = \lambda_s\qw_s,
\end{equation}
where $\qH_n\in \mathbb{C}^{N_\text{r} \times N_\text{t}}$ denotes the channel matrix for the $n$-th subcarrier ($n \in {1, \ldots, N_\text{c}}$), and $\lambda_s$ is the corresponding $s$-th largest eigenvalue. The precoding matrix is constructed as
\begin{equation}
\qW = [\qw_1, \qw_2, \dots, \qw_{N_\text{s}}]\in\bbC^{N_\text{t}\times N_\text{s}}.
\end{equation}

The 5G NR system uses a codebook-based approach to feedback the precoding matrix. Specifically, the Type II codebook introduced in 3GPP Release-15 \cite{3GPP_CSI_Feedback} employs Dual-Codebook technology based on the Type I codebook. This codebook supports up to two spatial streams and reconstructs the precoding matrix by combining up to four oversampled orthogonal discrete Fourier transform (DFT) beams. Although its performance approaches that of perfect feedback, the significant overhead reduces system efficiency. To address this challenge, we consider a DT-based approach to approximate RW scenarios.

\subsection{DT Framework}
The RW channel, denoted by $\mathcal{H}$, is influenced by three primary components:
(i)  {\bf Environmental Conditions} $\mathcal{E}$, including physical obstructions, atmospheric variations, reflections, and interference.
(ii) {\bf Antenna Effects} $\mathcal{A}$, determined by design factors such as gain, radiation patterns, and polarization.
(iii) {\bf Propagation Phenomena} $g(\cdot)$, which govern signal behaviors such as reflection, diffraction, and scattering. Thus, the RW channel is expressed as:
\begin{equation}
\mathcal{H} = g(\mathcal{E}, \mathcal{A}).
\end{equation}

Accurately modeling $\mathcal{E}$, $\mathcal{A}$, and $g(\cdot)$ is challenging. Thus, we approximate $\mathcal{H}$ with a virtual channel $\widetilde{\mathcal{H}}$ generated via advanced simulations:
\begin{itemize}
\item {\bf 3D Model $\widetilde{\mathcal{E}}$:} Detailed 3D models capture the positions, orientations, shapes, and materials of key components (BS, UE, reflectors, and scatterers), with material properties like conductivity and dielectric constants affecting signal propagation.

\item {\bf Antenna Properties $\widetilde{\mathcal{A}}$:} Antenna effects, including gain, phase, type, placement, and polarization, are modeled to simulate transmitter and receiver behavior accurately.

\item {\bf Ray Tracing $\tg(\cdot)$:} This deterministic method uses 3D geometry and material data to simulate propagation paths, accounting for transmission, reflection, scattering, and diffraction, and providing path gain, delay, and angular parameters.
\end{itemize}

By integrating all parameters from the ray tracing simulation, the virtual channel at the $n$-th subcarrier is modeled as:
\begin{equation}\label{eq:channel}
\widetilde{\qH}_n = \sum_{l=1}^{L_\text{h}} \qG(\qtheta^{\mathrm{AoD}}_l, \qtheta^{\mathrm{AoA}}_l)e^{-j2\pi \frac{n}{N_\text{c}}\tau_l},
\end{equation}
where $L_\text{h}$ is the total number of propagation paths; $\qtheta^{\mathrm{AoD}}_l$ and $\qtheta^{\mathrm{AoA}}_l$ are the angles of departure (AoD) and arrival (AoA) of the $l$-th path, respectively, both including azimuth and elevation components. The function $\qG(\cdot) \in \mathbb{C}^{N_\text{r} \times N_\text{t}}$ is the complex gain matrix determined by the AoD and AoA. The term $\tau_l$ is the propagation delay associated with the $l$-th path.

An accurately constructed virtual channel $\widetilde{\mathcal{H}}$ enables the approximation of the virtual precoding matrix $\widetilde{\qW}$, closely matching its RW counterpart. However, due to modeling limitations in reflections, diffractions, antenna patterns, and environmental noise, DTs cannot fully replicate RW channels. To address this, we propose a hybrid approach that integrates DL with DT data. As shown in Fig. \ref{fig:Key_Point_NN}, the model is initially trained on synthetic DT data and subsequently fine-tuned using limited RW data, thereby enhancing accuracy and robustness.

\subsection{CSI Feedback with DTs}
To enable implicit feedback, we adopt the DL-based autoencoder EVCsiNet \cite{EVCsiNet}, which is designed for reduced computational complexity and latency. As illustrated in Fig.~\ref{fig:Key_Point_NN}, neural networks for encoding and decoding are deployed at the UE and BS, respectively. The complex-valued input precoding matrix is preprocessed by separating its real and imaginary parts. The encoder and decoder, parameterized by $\Theta_{\mathrm{E}}$ and $\Theta_{\mathrm{D}}$, are denoted as $f_\text{e}(\cdot; \Theta_{\mathrm{E}})$ and $f_\text{d}(\cdot; \Theta_{\mathrm{D}})$, respectively. The overall autoencoder, $f_\text{a}(\cdot; \Theta)$ with $\Theta = (\Theta_{\mathrm{E}}, \Theta_{\mathrm{D}})$, is expressed as:
\begin{equation}
f_\text{a}(\qW; \Theta)  = f_\text{d}(Q(f_\text{e}(\qW; \Theta_E)); \Theta_D),
\end{equation}
where $Q(\cdot)$ denotes the quantizer. The encoder compresses the virtual precoding matrix $\qW$ into a $V$-dimensional vector that is quantized into a codeword using $B$ bits, while the decoder reconstructs $\qW$ from the codeword.

The DL optimization objective using virtual channels $\widetilde{\mathcal{H}}$ is given by
\begin{equation}\label{eq:DL_vir}
\dot{\Theta} = \left( \dot{\Theta}_{\mathrm{E}}, \dot{\Theta}_{\mathrm{D}}\right)  = \argmin_{\Theta} \mathbb{E}_{\widetilde{\mathcal{H}}}\left\lbrace L\left( \widetilde{\qW}, f_\text{a}(\widetilde{\qW}; \Theta)\right) \right\rbrace,
\end{equation}
where $L(\cdot)$ is the mean square error (MSE) loss function, and $\mathbb{E}_{\widetilde{\mathcal{H}}}\{\cdot\}$ denotes the average over $\widetilde{\mathcal{H}}$.
To bridge the gap between virtual and RW environments, OL is employed to fine-tune the model. Due to constraints such as battery life, computational load, and limited resources at the UE, OL is primarily conducted at the BS by fine-tuning only the decoder $f_\text{d}(\cdot)$. The fine-tuning objective is
\begin{equation}\label{eq:tranL}
\ddot{\Theta}_{\mathrm{D}}= \argmin_{\Theta_{\mathrm{D}}} \mathbb{E}_{\mathcal{H}}\left\lbrace L\left( \qW, f_\text{a}(\qW; \dot{\Theta})\right) \right\rbrace,
\end{equation}
where $\dot{\Theta}$ is initialized from virtual channel training. Notably, this OL approach introduces minimal communication and computation overhead, as the encoder remains fixed and only the decoder is retrained at the BS.

\begin{figure*}
    \begin{center}
        \resizebox{7in}{!}{
            \includegraphics*{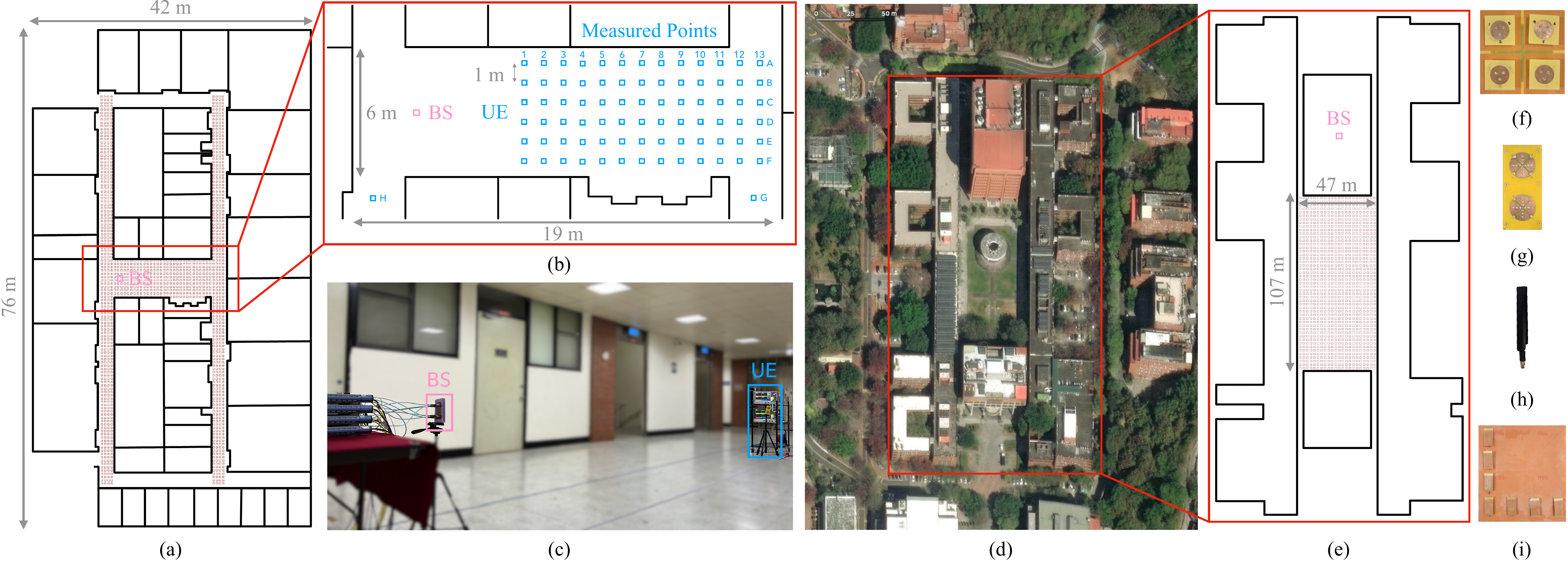} }
            \caption{(a) Simulated indoor corridor at NSYSU using Wireless InSite\textsuperscript{\textregistered}. (b) BS and UE RW measurement point locations. (c) Experimental measurement scenario. (d) Satellite image of the NSYSU campus. (e) Simulated outdoor scenario at the NSYSU campus using Wireless InSite\textsuperscript{\textregistered}. (f) 15 cm x 15 cm patch antenna at the BS for the CSI feedback experiment. (g) 7 cm x 15 cm patch antenna at the UE for the CSI feedback experiment. (h) 11.4 cm dipole antenna at the BS for the channel fidelity experiment. (i) 15.5 cm x 17 cm patch antenna at the UE for the channel fidelity experiment.}
        \label{fig:Configuration} 
        \vspace{-0.35cm}
    \end{center}
\end{figure*}

\section{Experimental Setup}

\subsection{RW Scenario}
RW measurements were conducted in an indoor corridor at National Sun Yat-sen University (NSYSU) as depicted in Fig.~\ref{fig:Configuration}(a)-(c). The corridor measures approximately 19.7 m in length and 5.93 m in width, with a ceiling height of 2.8 m. The BS, an SGT100A RF vector signal generator with an 8-patch antenna (Fig.~\ref{fig:Configuration}(f)), is positioned at the left-center of the area, while the receiver is an RTP084 digital oscilloscope with an 8-patch antenna (Fig.~\ref{fig:Configuration}(g)). Data were collected at 78 marked points, with 200 measurements per point.
The communication system follows the 5G NR standard \cite{3GPP_Waveform} and operates at 3.8 GHz using an OFDM waveform with a 100 MHz channel bandwidth, 60 kHz subcarrier spacing, and 1,620 subcarriers. The BS transmits two spatial streams ($N_\text{s}=2$), each delivering $8.64 \times 10^7$ resource elements per second with 1024-QAM modulation. An expectation propagation-based (EP) MIMO detector is used to evaluate the throughput in this scenario.

\subsection{Virtual Channel in DT}
The virtual channel $\widetilde{\mathcal{H}}$ is generated using Wireless InSite\textsuperscript{\textregistered} ray-tracing simulations that replicate the RW environment with a detailed 3D model, accurately modeling reflective surfaces, BS/UE positions, and orientations. The UE movement area (red points in Fig.~\ref{fig:Configuration}(a)) is discretized at 0.5-meter intervals, with each UE orientation randomly chosen from 100 possible directions. Antenna properties, measured in an anechoic chamber, ensure realistic simulation results.
Based on the extracted propagation parameters, the virtual channels are computed as follows.
Using the ray-traced propagation parameters, the channel between the BS and UEs is computed according to \eqref{eq:channel}, and the corresponding virtual precoding matrix is derived via \eqref{eq:eig_de}. The simulation waveform design matches that of the RW scenario, thereby establishing the indoor DT dataset. 
To evaluate the robustness of the DL-based CSI feedback model, additional virtual channels are generated under various conditions, including different environmental and antenna configurations. An outdoor virtual scenario replicates the NSYSU campus (Fig.~\ref{fig:Configuration}(d)-(e)), with the BS positioned 50 meters above a building and UEs distributed at 2-meter intervals across the campus square. This simulation yields the outdoor DT dataset.

\subsection{Fidelity of Virtual Channel}
To assess the fidelity of the virtual channel, we compare the simulated AoA from the DT with RW AoAs measured in an indoor corridor at NSYSU (Fig.~\ref{fig:Configuration}(a)-(b)). Because the standard antennas (Fig.~\ref{fig:Configuration}(f)-(g)) are suboptimal for AoA extraction, a vertical dipole at the BS (Fig.~\ref{fig:Configuration}(h)) and a 7-patch array with half-wavelength spacing at the receiver (Fig.~\ref{fig:Configuration}(i)) were employed instead. The AoA is computed from the phase difference of the received plane wave.
The DT AoA is represented by the 3D direction vector
$\qu(\phi,\theta)=
\big[\cos(\phi)\cos(\theta), \
 \cos(\phi)\sin(\theta), \
 \sin(\phi) \big]^\mathrm{T}$,
where $\theta$ and $\phi$ denote the azimuth and elevation angles, respectively.  The estimated AoA is similarly represented with $(\hat\phi,\hat\theta)$, and the similarity between the two is measured via the inner product
\begin{equation}
\eta= \big\langle \qu(\phi,\theta), \ \qu(\hat\phi,\hat\theta) \big\rangle, \, -1 \le \eta \le 1.
\end{equation}
Five measurement points were selected (A4, D4, F4, G, and H; see Fig.~\ref{fig:Configuration}(b)), with A4, D4, and F4 in strong signal areas and G and H in weaker regions. Table~\ref{tab:Precoder_Similarity} lists the similarity values for the primary (strongest gain) and secondary (second strongest) paths, indicating high similarity between the DT and RW propagation paths.

\begin{table}[]
\centering \footnotesize
\caption{Similarity of propagation paths between the DT and the RW.}
\label{tab:Precoder_Similarity}
\begin{tabular}{c ccccc}
    \toprule
        \textbf{Position}              & A4    & D4   & F4    & G     & H     \\ \cmidrule{1-6}
        \textbf{Primary Path}      & 0.98 & 1.00 & 0.98 & 0.95 & 0.98 \\
        \textbf{Secondary Path} & 0.93 & 0.85 & 0.99 & 0.99 & 0.89 \\
    \bottomrule
\end{tabular} 
\end{table}

\begin{figure}
    \begin{center}
        \resizebox{3in}{!}{
            \includegraphics*{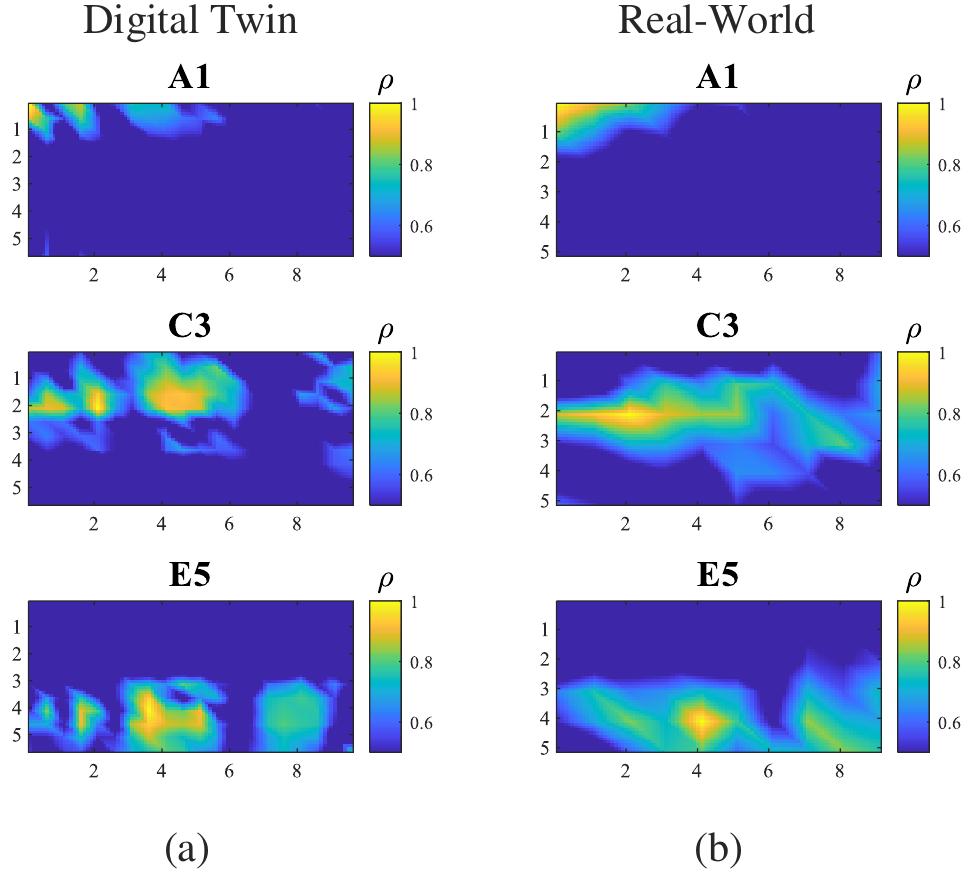} }
        \caption{Heatmaps of precoder cosine similarity in (a) DT and (b) RW datasets.}
        \label{fig:Precoder_Similarity}
        \vspace{-0.35cm}
    \end{center}
\end{figure}

Since the precoder $\qW$ is the main input for EVCsiNet, we further compare the scenario similarity using the precoder cosine similarity. Fig.~\ref{fig:Precoder_Similarity} shows that both the DT and RW datasets exhibit high similarity for positions aligned in the same direction relative to the BS. These findings, from both AoA and precoder perspectives, validate that the DT dataset effectively reflects RW conditions.

While the DT dataset closely approximates RW conditions, some discrepancies remain due to practical antenna effects. Consequently, OL is necessary. For OL, as in \eqref{eq:tranL}, RW precoding matrix data $\qW$ is required. To balance performance and data collection costs, 30\% of the RW measurement points in Fig.~\ref{fig:Configuration}(b) were randomly selected. Furthermore, to maximize the information extracted from limited measurements, the 1,620 subcarriers were grouped into 60 subbands, each containing 27 consecutive subcarriers. This configuration enables each measurement point to generate 60 distinct precoding matrices for OL, thereby enhancing model adaptation with minimal data collection.

\section{Evaluations and Discussions}
\label{sec_DT_Performance}

We evaluate three EVCsiNet variants: \textbf{Indoor EVCsiNet}, trained on an indoor DT dataset; \textbf{Outdoor EVCsiNet}, trained on an outdoor DT dataset; and \textbf{CDL EVCsiNet}, the baseline trained on the standardized Cluster Delay Line (CDL) model from 3GPP TR 38.901. All adopt a compression ratio of 0.5, with 16-length codewords quantized to 2 bits per element, resulting in a 32-bit feedback overhead. Reconstruction performance is measured by cosine similarity (\( \rho \)).

\subsection{Performance of the DT-Trained Model in RW Scenarios}

Table~\ref{tab:Rho_Comparison} compares precoding matrix reconstruction performance of various CSI feedback methods (fourth column). Among them, the Type II method with 4 beams achieves the highest reconstruction performance ($\rho=0.96$) but requires the most feedback bits (80 bits). Notably, Indoor EVCsiNet with OL achieves reconstruction performance comparable to the Type II method with 3 beams (i.e., $\rho=0.82$ vs. $0.86$) while significantly reducing the overhead (32 bits vs. 58 bits).

\begin{table}[!t]
\centering \footnotesize
\caption{Reconstruction Performance Across Feedback Methods.}
\label{tab:Rho_Comparison}
\renewcommand{\arraystretch}{1.3}
\begin{tabular}{c c c c c}
\toprule
\multirow{2}{*}{\makecell[c]{\textbf{Feedback}\\\textbf{Method}}} & \multirow{2}{*}{\makecell[c]{\textbf{Training}\\\textbf{Environment}}} &  \multirow{2}{*}{\textbf{Bits}} & \multicolumn{2}{c}{$\rho$} \\ 
& & & \textbf{RW} & \textbf{Indoor DT} \\
\midrule
\multirow{3}{*}{\makecell[c]{EVCsiNet\\(w/o / w/ OL)}}
& Indoor DT    & 32 & 0.67 / 0.82 & 0.93 / ~\textemdash~ \\
& Outdoor DT & 32 & 0.64 / 0.75 & 0.77 / 0.85                   \\
& CDL            & 32 & 0.44 / 0.74 & 0.67 / 0.77                   \\
\midrule
\multirow{3}{*}{\makecell[c]{Type II\\Codebook}}
& $N=2$ beams  & 41         & 0.72 & 0.70 \\ 
& $N=3$ beams  & 58         & 0.86 & 0.84 \\ 
& $N=4$ beams  & 80         & 0.96 & 0.96 \\
\bottomrule
\end{tabular}
 \vspace{-0.5cm}
\end{table}

Although directly applying the Indoor EVCsiNet to RW scenarios yields suboptimal performance, it still outperforms both the Outdoor and CDL variants. With OL, the Indoor EVCsiNet further improves and maintains a clear advantage over the other two. These results indicate that DT-based pretraining offers better initialization by reducing domain mismatch and significantly enhances the effectiveness of OL. In contrast, CDL-based pretraining followed by OL remains insufficient for achieving robust CSI reconstruction in RW conditions.

\begin{figure}
    \begin{center}
        \resizebox{3.25in}{!}{
            \includegraphics*{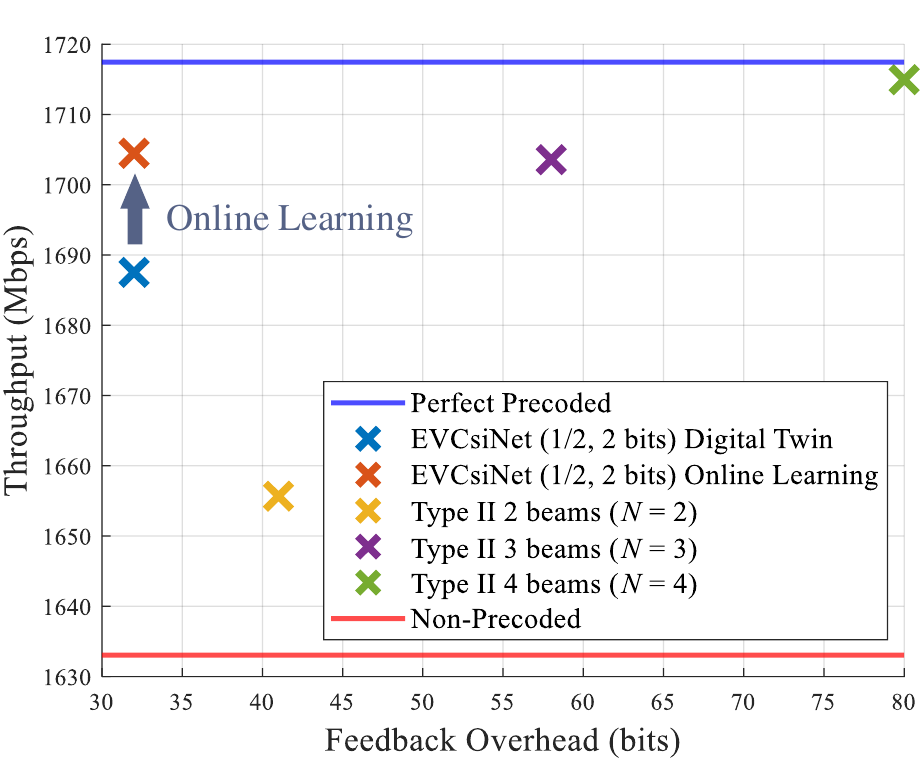} }
        \caption{RW throughput comparison across different feedback methods.}
        \label{fig:Throughput}
        \vspace{-0.35cm}
    \end{center}
\end{figure}

Fig.~\ref{fig:Throughput} compares the feedback overhead and throughput of the Indoor EVCsiNet with other CSI feedback methods in RW scenarios. Applying OL yields a 17 Mbps increase in throughput for the Indoor EVCsiNet, achieving performance comparable to the Type II method with 3 beams. Interestingly, the Indoor EVCsiNet outperforms the Type II method with 2 beams in terms of throughput despite exhibiting a lower similarity. This discrepancy is attributed to the unbalanced performance of the Type II method with 2 beams; with only 2 beams, the precoder corresponding to each of the largest eigenvalues is not accurately reconstructed, which hampers the BS's ability to transmit signals through the dominant eigenchannels effectively. In contrast, EVCsiNet reconstructs each precoder with balanced performance, avoiding this issue.

\subsection{Robustness of the DL Model}

Table~\ref{tab:Rho_Comparison} shows that the Indoor EVCsiNet suffers performance degradation in RW scenarios due to mismatches in environment $\mathcal{E}$ and antenna effects $\mathcal{A}$. To assess model robustness and isolate the impact of $\mathcal{A}$, we evaluate all models on the indoor DT dataset with fixed antenna settings. The last column reports the similarity metric $\rho$ for each feedback method. The Indoor EVCsiNet achieves the best performance ($\rho = 0.93$), outperforming the Outdoor EVCsiNet ($\rho = 0.77$) and CDL EVCsiNet ($\rho = 0.67$), while using fewer feedback bits. It also approaches the 4-beam Type II codebook ($\rho = 0.96$), which requires significantly higher overhead. These results show that the Indoor EVCsiNet benefits from matched training conditions, whereas the CDL model struggles to generalize to the DT scenario.

Furthermore, incorporating 5\% of the indoor DT dataset into OL with the Outdoor EVCsiNet improved its reconstruction similarity to 0.85, achieving performance comparable to the Type II method with 3 beams. Nevertheless, the Outdoor EVCsiNet underperformed relative to the Indoor EVCsiNet, underscoring the importance of matching environmental conditions during offline learning. While OL can mitigate environmental mismatches, it cannot fully compensate for them.

We next assess the impact of antenna effects $\mathcal{A}$ on Indoor EVCsiNet. Table~\ref{tab:Antenna_Comparison} shows that UE antenna changes have minimal impact, while BS antenna variations significantly degrade performance, more than environmental changes, due to the precoder’s sensitivity to BS directionality. This degradation occurs because the precoder highly depends on the BS antenna's directionality. While environmental changes prevent the DL model from operating under optimal conditions, any alteration in the BS antenna pattern directly affects its directional characteristics, leading to a more substantial performance decline.

Incorporating 5\% of new BS antenna data into OL restores performance to $\rho = 0.84$, demonstrating that OL mitigates antenna variations.
Although the detailed results are omitted for brevity, further evaluations reveal that performance remains stable under time-varying interference and user mobility, suggesting that these effects do not need to be explicitly modeled in the DT.
Moreover, while EVCsiNet requires retraining under the corresponding DT as the number of BS antennas increases, the OL method adapts using only a small amount of RW data.
Furthermore, increasing the feedback bits provides only marginal gains, as most of the precoding matrix can already be recovered at lower feedback rates, while OL enables robust CSI reconstruction under distribution mismatch.

\begin{table} 
\centering \footnotesize
\renewcommand{\arraystretch}{1.3}
\caption{Antenna Pattern Effects on Reconstruction Performance.}
\label{tab:Antenna_Comparison}
\begin{tabular}{ccc}
\toprule
\textbf{Feedback Method} & \textbf{Antenna Pattern}  & $\rho$ \\
\midrule
\multirow{4}{*}{EVCsiNet}
& Original                                 & 0.93 \\
& Change UE Pattern              & 0.94 \\
& Change BS Pattern              & 0.57 \\
& Change BS Pattern with OL & 0.84 \\
\bottomrule
\end{tabular}
\vspace{-0.3cm}
\end{table}

\section{Conclusions}
This paper investigated a DL-based CSI feedback model trained in a DT environment and tested in RW scenarios. The results indicate that integrating minimal RW data via OL is essential to mitigate performance degradation arising from discrepancies between DT and RW datasets. Furthermore, environmental conditions and BS antenna characteristics significantly impact DL model training, highlighting the importance of a dedicated DT even when OL is employed.

\bibliographystyle{IEEEtran}
\bibliography{References}

\end{document}